\documentclass[11pt]{article}
\usepackage{amssymb}
\usepackage{amsthm}
\usepackage{amsmath,amsfonts,amssymb}

\usepackage[mathscr]{eucal}
\usepackage{exscale}
\usepackage{natbib}
\usepackage{bm}
\usepackage{eqlist}
\usepackage[dvipsnames]{color}
\usepackage[Lenny]{fncychap}
\usepackage{multirow}
\usepackage{graphicx}
\usepackage{algpseudocode}
\usepackage{algorithm}
\usepackage{caption}
\usepackage{hyperref}
\usepackage{framed}
\usepackage{color}
\usepackage{booktabs}
\usepackage{makecell}
\usepackage{array}
\usepackage{float}
\usepackage{placeins}
\usepackage{tabularx}
\usepackage{longtable}
\usepackage[table]{xcolor} % 引入xcolor宏包

\usepackage{framed}
\usepackage{algpseudocode}

\usepackage{tabu}
\usepackage{tikz}

\usepackage[utf8]{inputenc} % this is needed for umlauts
\usepackage[english]{babel} % this is needed for umlauts
\usepackage[T1]{fontenc}    % this is needed for correct output of umlauts in pdf
\usepackage{amssymb,amsmath,amsfonts} % nice math rendering
\usepackage{braket} % needed for \Set
\usepackage{caption}
\usepackage{algorithm}

\hypersetup{
    colorlinks=true,
    linkcolor=blue,
    citecolor=blue,
    citebordercolor={1 1 0},
}

\renewcommand{\theequation}{\thesection.\arabic{equation}}

\usepackage{authblk}

\title{Inferring ghost cities on the globe in newly developed urban areas based on urban vitality with multi-source data}

\author{
Yecheng Zhang\textsuperscript{a},
Tangqi Tu\textsuperscript{a},
Ying Long\textsuperscript{a,b} \thanks{Corresponding author, e-mail: ylong@tsinghua.edu.cn.} \\
\small \textsuperscript{a} \textit{School of Architecture, Tsinghua University, Beijing, 100084, China} \\
\small \textsuperscript{b} \textit{Hang Lung Center for Real Estate, Key Laboratory of Ecological Planning $\&$ Green Building, Ministry of Education, Tsinghua University, Beijing, 100084, China}
}
\date{\today} % 显示当前日期

\begin{document}

\maketitle

\begin{abstract}
Due to rapid urbanization over the past 20 years, many newly developed areas have lagged in socio-economic maturity, creating an imbalance with older cities and leading to the rise of "ghost cities". However, due to the complexity of socio-economic factors, no global studies have measured this phenomenon. We propose a unified framework based on urban vitality theory and multi-source data, validated by various data sources. We derived 8841 natural cities globally with an area over 5 km², and divided each into new urban areas (developed after 2005) and old urban areas (developed before 2005). Urban vitality was gauged using the density of road networks, points of interest (POIs), and population density with 1 km resolution across morphological, functional, and social dimensions. By comparing urban vitality in new and old urban areas, we quantify the ghost cities index (GCI) globally using the theory of urban vitality for the first time. The results reveal that the vitality of new urban areas is 7.69\% that of old ones. The top 5\% (442) of cities were designated as ghost cities, a finding mirrored by news media and other research. This study sheds light on strategies for sustainable global urbanization, crucial for the United Nations' Sustainable Development Goals.

\end{abstract}


\renewcommand{\theequation}{\arabic{equation}} % 重新定义公式编号
\setcounter{equation}{0} % 重置公式计数器
\section{Introduction}\label{secIntro}
\subsection{Research background}
Due to urbanization in many countries, rapid urban land expansion has become a global phenomenon \citep{LI2020, LU2020, WA2022, LY2023, ZZ2024}. Especially in emerging economies and developing countries, numerous new urban areas have been built up. Yet, it is crucial to emphasize that the rate of urban socioeconomic progress has not paralleled this land expansion. Many new urban areas are thus inadequately developed at a low socioeconomic level, resulting in abundant vacancies in these new urban areas described as so-called "ghost cities" \citep{JN2017}. 'Ghost cities' traditionally describe urban areas abandoned due to factors like resource depletion or market shifts \citep{CZ2013, RO2011}. However, recent studies, especially by \citet{XU2022}, highlight a shift in understanding, emphasizing newly built but uninhabited areas, especially in developing regions. The United Nations' Sustainable Development Goal 11 (SDG11) is to make cities inclusive, safe, resilient, and sustainable \citep{UN2015}, and ghost cities are always generally considered to have negative impacts on developing sustainable cities \citep{BA2016, WH2024, ZZ2022}. To cope with ghost cities and achieve SDG11, it is essential to explore the existence of ghost cities and identify their characteristics on a global scale.

Understanding the definition of ghost cities is foundational to their identification. Some scholars have defined this term from various perspectives. The concept of “ghost cities” or “ghost estates” was originally proposed to describe vacant housing in Ireland \citep{OC2014}, which means the original ghost cities were once inhabited but became empty after the people left. However, with the rapid growth of cities globally in recent times, the term “ghost city” has been redefined from a new perspective, beginning to refer to the phenomenon of high vacancy rates, shrinking commercial activities, and sparse population in newly built urban areas due to over-development and improper planning \citep{WI2019}. \citet{SH2015} describes a “ghost city” as “a new development that is running at significantly under capacity, a place with drastically fewer people and businesses than there is an available space for”. Moreover, he elaborates on some phenomena that have appeared in Chinese cities as a typical warning of today's ghost cities. With the continuous development of data and technology, ghost cities in new urban areas have evolved in their definitions. These “ghost cities” in new urban areas generated from urban expansion have drawn concerns widely from scholars, social media, and governments \citep{BA2016, ZH2017, JI2015}. Therefore, this study primarily focuses on ghost cities from the perspective of the inadequate operating level of newly built-up urban land.

\subsection{Literature review}
Strictly identifying a city as a 'ghost city' is complex, especially on a global scale. Some researchers explored identifying and evaluating ghost cities on a local scale. In these studies, ghost cities are mainly identified and evaluated based on population \citep{ZH2022, LE2019}, firms \citep{DO2021}, nighttime satellite image data \citep{SHI2020, ZH2019, ZH2017, ZH2022}, and amenities \citep{WI2019}. On the one hand, these local scale analyses based on population or built environment indicators identified ghost cities defined by thresholds chosen without considering the consistency across geographical regions. That’s to say such a threshold is usually suitable for a local homogeneous geographical scale, and there are huge differences in urban construction between different geographical regions in the world. On the other hand, previous studies have revealed that ghost cities have been existing in the new urban areas of some specific cities or some major countries like the United States and China. Nevertheless, whether ghost cities exist widely in the world and how to identify them at the global scale remain in question, which is important for global sustainable urbanization, especially in many developing countries experiencing rapid urbanization. It is thus necessary to establish an effective framework to identify global ghost cities based on related theories at a more sophisticated level with multi-source data.

Furthermore, while a unified standard for defining ghost cities remains elusive \citep{JN2017, SHI2020, GZ2024}, their hallmark of low spatial vitality in new urban regions is evident \citep{GE2018}. This study leverages the urban vitality theory to identify global ghost cities by comparing the vitality of newer urban areas with that of older ones. \citet{LY1984} posits urban vitality as a metric of urban space quality, crucial for settlements that cater to human needs and ensure species survival. When assessing this vitality, numerous researchers utilize indicators related to 'urban form', 'life place', and 'human activity', concepts introduced by both \citet{JA1961} and \citet{LY1984}. These indicators capture critical urban design elements \citep{LH2019}. Urban vitality often parallels a high 'quality of life'. Jacobs underscores the interplay between human activity and life space, suggesting that a city's vitality reflects its diverse life. Building on these theories, this study decomposes urban vitality into three key elements for calculation using accessible data: the morphological, functional, and social dimensions of urban vitality. As highlighted by \citet{JN2017}, urban vitality forms the foundation for a quality urban life, derived from good urban form, developed urban functions, and sufficient urban activities. In this way, the relative vitality inadequacy of new urban areas can be presented when compared with corresponding old urban areas, which can provide the quantitative foundation for identifying ghost cities.

While this vitality theory established the basis for identifying ghost cities, the current section delves deeper into the specifics, informed by extensive literature research. First, the morphological dimension of urban vitality, represented by "urban form", is influenced by factors such as development density, providing the physical environment for urban vitality \citep{KR2017, YE2018, AK2022, WU2018, CH2022, DZ2022}. Second, the functional dimension, or "life place", is influenced by economic development, with higher levels leading to diverse functions that stimulate urban utilization \citep{HU2020, CH2022, YU2021}. Third, the social dimension, or "human activity", is reflected in the spatial concentration of people, with increased social interactions contributing to urban vitality \citep{ZH2012, GU2022, SU2013}. These dimensions are measured by different indicators. Road density measures the morphological dimension, as denser urban roads increase accessibility and potential for urban vitality \citep{PA2022, YU2021}. The functional dimension is measured by Points of Interests (POIs), which provide detailed types of functions and precise locations \citep{YU2021, WI2019, DM2018, GV2022, DO2021, JI2015}. The social dimension is best represented by population distribution data, as denser populations tend to have more interactions, contributing to urban vibrancy \citep{ZH2022, LE2019, DO2021, SHI2020, ZH2019, ZH2017, GW2016, JN2017}. Thus, this study uses road density, POIs density, and population density to represent the morphological, functional, and social dimensions, respectively.

\subsection{Problem statement and objectives}
Given the paramount importance of identifying global ghost cities for policy formulation, urban planning, and sustainable urbanization, there's an urgent need to establish a quantifiable and universally comparable framework. Although many studies have been done to predict ghost cities in a local area through indicators such as population distribution and built environment, these methods are not extensible on a global scale. Therefore, the main innovation of this study is constructing an effective framework to identify global ghost cities based on urban vitality differences with more sophisticated methods and more accessible data. In this article, we derive 8841 natural cities through the data of urban boundaries in 2005 and 2018 \citep{LI2020, XU2021} at first, which we refer to as "natural cities" throughout this article. By dividing natural cities into new urban areas (developed after 2005) and old urban areas (developed before 2005), we borrow the urban vitality theory to measure the vitality gap between the new and old urban areas in the natural cities with 1 km resolution, calculating the ghost city index (GCI) and infer the distribution of ghost cities in newly developed urban areas. Referring to this, we also analyzed the characteristics of global ghost cities in various countries and regions. Subsequently, we verified our identified results in different scales using Google search engines and related research, which showed consistent results. At the same time, in order to understand the spatial distribution pattern of the GCI in this paper, the demographic factors have been considered in the research, and verified our results through the actual population growth of natural cities. The ghost city index we gave illustrated a strong potential to overcome the difference in urban scales across cities on the globe. Using this unified framework, we can discuss the possibility of the widespread distribution of new urban areas in the natural cities on the globe and observe differences in scale at the intercontinental, national, and urban levels in a quantitative manner. Our result can provide a scientific reference for global urban planning to achieve sustainable urbanization.

\section{Methodology}\label{secIntro}
\subsection{Methodology Framework}
We used a unified spatial analysis unit to aggregate data from different dimensions to calculate the vitality index of new and old urban areas for each natural city and obtained the Ghost City Index (GCI) for each natural city using the corresponding formula. The three dimensions of measuring urban vitality include morphological, functional, and social dimensions. Based on the literature review mentioned above, we selected road network density, POI density, and population density to represent the three dimensions, respectively. These indicators can be further supported by more accurate and open data in the future to analyze more detailed scales. Our method includes three steps: 

\begin{enumerate}
    \item \textbf{Data preprocessing:} Based on interpreted remote sensing products \citep{LI2020, XU2021} GHUB and GUB data, globally comparable natural city expansion data were obtained through geospatial matching (see section 2.3 Data acquisition and 2.4 Global urban areas). At the same time, the global population, POI, and road network data are collected and unified to the 1-kilometer grid. With further unification of urban definitions by the United Nations, urban boundary data will become more available \citep{LI2024}.
    \item \textbf{Calculation of the Ghost City Index:} Indicators for the three dimensions and urban vitality of new and old urban areas were calculated, and then the Ghost City Index for each natural city was derived and analyzed at global, intercontinental, national, and urban scales (see section 2.2 Calculation of GCI).
    \item \textbf{Validation of calculation results:} Validation was conducted at two levels. First, by scraping Google keyword searches, we obtained the popularity rankings of ghost city search terms for countries and regions globally and compared them with our results. Second, we calculated population and land expansion for all natural cities across different years to verify whether the inference of ghost cities based on the demographic factor from new and old urban areas is reasonable (see section 3.5 Benchmarking our results from news media and related research and 3.6 Understanding the distribution of GCI from different perspectives).
\end{enumerate}

% 插入图片
\begin{figure}[h!] % 使用 [!htbp] 选项控制图片位置
\centering
\includegraphics[width=0.90\textwidth]{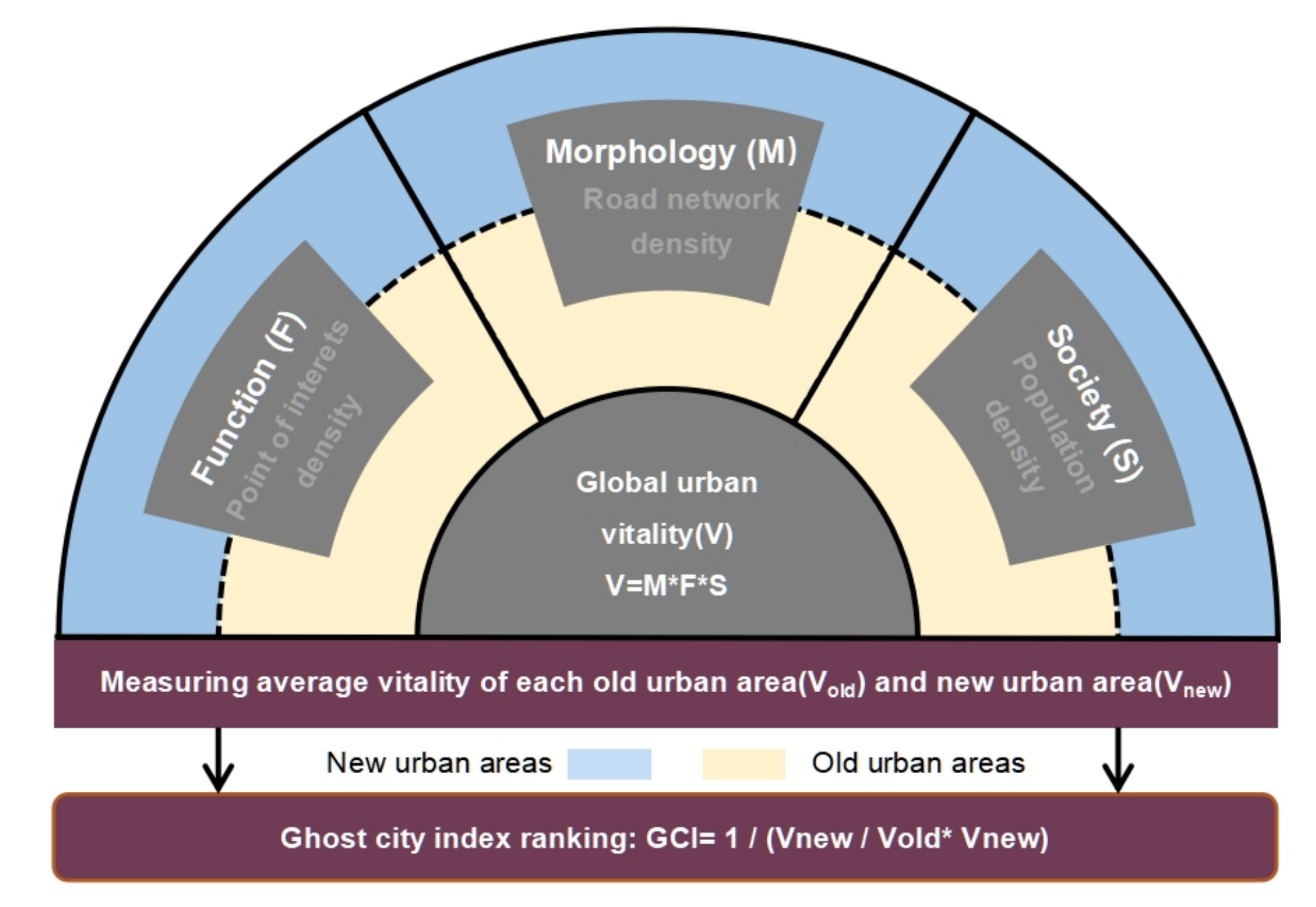} 
\caption{\textbf{Methodological framework of this study.} The degree to which a city belongs to a ghost city is related to vitality in new urban areas and the difference in vitality between old and new urban areas. The larger the difference and the lower the vitality in the new residential projects, the more likely the city to be a ghost city.}
\label{fig:stream}
\end{figure}
\FloatBarrier % 确保图片在此之前出现

\subsection{Calculation of GCI}
In terms of measuring the vitality of urban areas, a straightforward quantitative method is employed to measure the vitality of all the urban areas on the globe \citep{JN2017}. First, the density distribution layers of road junctions, POIs, and population are calculated and generated by the software ArcGIS. Each output raster is set with a cell size of 1 km to maintain the consistency of these three layers covering the entire world. The second step is to measure the average value of each layer within each urban area. Finally, the vitality of each urban area is measured by the equation below:
\[
V = M \times F \times S
\]
where \( V \) is the vitality; \( M \) is the density of road network as the indicator from the morphological dimension; \( F \) is the density of POIs as the indicator from the functional dimension; \( S \) is the population density as the indicator from the social dimension. This equation indicates that an urban area with denser roads, more urban functions, and more human activities tends to have a higher vitality level.

Based on the results of measuring the vitality of global urban areas, the ghost city index of new urban areas can be calculated for all the cities on the globe by combining the vitality values of the new and old urban areas in each city. The equation that can be used is \citep{JN2017}:
\[
GCI = \frac{1}{{\left( \frac{V_{\text{new}}}{V_{\text{old}}} \right) \times V_{\text{new}}}}
\]
where \( GCI \) is the ghost city index of the new urban area in a city; \( V_{\text{old}} \) is the average vitality of all the grids in the old urban area in a city; \( V_{\text{new}} \) is the average vitality of all the grids in the new urban area in a city. According to this equation, the ghost city level of the new urban area that a city belongs to is associated with the vitality of the new urban area and the vitality difference between new and old urban areas in the city. Suppose the vitality of the new urban area is lower or the vitality difference between new and old urban areas is larger. In that case, this index will be higher and the new urban area of the city is more likely to be a so-called “ghost city” \citep{JN2017}. All the cities can be then ranked based on this index to identify ghost cities by higher index values.

\subsection{Data acquisition}
Multi-source data were acquired to support this study. The data used for the analysis include global urban boundaries, density of road network, density of points of interest (POIs), and population density. The basic information about the resolution, temporal domain, type, and source of the data is shown in Table 1. The detailed description of the data is as follows. The road data and POIs data used in this paper are all from OpenStreetMap (OSM). As an open-source geographic database that is considered as the infrastructure network is nearing completion, OSM provides the only publicly licensed data sources such as geospatial roads, POIs, and building footprints at the global level. Many studies have tried to evaluate the data integrity and usability of OSM, including different scales such as cities, regions, and the world, proving the availability of road data and POIs data, especially now \citep{KL2023, PI2023, OO2023, BI2023, HE2023}.

\begin{table}[ht]
\centering
\caption{Basic information of the data used in this study}
\small % Reduce font size
\resizebox{\textwidth}{!}{% Adjust table size to fit within text width
\begin{tabular}{|m{4.2cm}|m{3.2cm}|m{1.5cm}|m{6.5cm}|}
\hline
\textbf{Name} & \textbf{Temporal domain} & \textbf{Type} & \textbf{Source} \\
\hline
Global urban boundaries & 2018, 2005 & Polygon & Li et al., 2020a; Xu et al., 2021; \url{http://data.ess.tsinghua.edu.cn/} \newline \url{https://doi.org/10.5281/zenodo.5168383} \\
\hline
Density of road network & 2022 & Polyline & OpenStreetMap \newline \url{http://download.geofabrik.de/} \\
\hline
Density of POIs & 2022 & Point & OpenStreetMap \newline \url{http://download.geofabrik.de/} \\
\hline
Population density & 2020/2005 & Raster / 1 km resolution & WorldPop unconstrained global population grids \newline \url{https://www.worldpop.org/} \\
\hline
National boundaries & 2022 & Polygon & OpenStreetMap \newline \url{http://download.geofabrik.de/} \\
\hline
Administrative boundaries & 2022 & Polygon & Database of Global Administrative Areas (GADM) \newline \url{https://gadm.org/} \\
\hline
\end{tabular}%
}
\end{table}

\subsection{Global urban areas}
The spatial distribution of global urban areas was determined based on data from two related studies \citep{XU2021, LI2020} which used global hierarchical urban boundaries (GHUB) and global urban boundaries (GUB) from 2018 and 2005, respectively. Both datasets employed the widely used definition of urban boundary, which was identified using the same 30m global artificial impervious area (GAIA) data with a mean overall accuracy higher than 90\% \citep{GO2020}. New urban areas in this study refer to those developed between 2005 and 2018, while old urban areas refer to those developed before or in 2005. Due to the complexity of urban land changes over time, we adopted a "converting time to space" approach to compare urban areas built in different years within the same city (Fig 2). We used the urban boundaries in 2018 as the benchmark spatial analysis units, allowing for the division of global new urban areas (GNUA) and global old urban areas (GOUA) within a city for year-to-year comparisons based on observed built-up years.

We chose GHUB instead of GUB in 2018 for two primary reasons. Firstly, GHUB data displays better consistency in urban boundary levels with lower fragmentation characteristics, negating the need to screen for natural city thresholds again. Additionally, GHUB data offers smoother, continuous boundary contours and a more comprehensive analysis of urban components compared to GUB. Our research solely evaluates urban land expansion, eliminating data on the physical shrinkage of urban land. As per this principle, we utilized ArcGIS software to connect, fuse, and remove new city data from 2005 without the old city based on the 2018 city boundary data. Fig.8 represents only typical expansion situations of urban land, excluding shrinkage circumstances.

Our analysis considered global urban boundary delineations for 2005 and 2018, with 9,862 cities initially designated. After data cleaning (appendix), 8,841 natural cities were retained as benchmarks, totaling 750,800 square kilometers of area in 2018. Compared to other datasets, our dataset is more accurate and efficient in representing global urban boundaries' spatial details. For example, our dataset highlights more urban fringe areas detail than night-time light-derived data, while maintaining similar urban dynamic consistency \citep{ZH2018}. The dataset aligns well with human-interpreted data in terms of overall spatial distribution and spatial details around urban fringe areas, reducing intensive human labor \citep{WA2012}. Our dataset reflects the actual spatial extent and urban dynamics for large and small cities globally, making it internationally comparable \citep{XU2021}.

% 插入图片
\begin{figure}[h!] % 使用 [!htbp] 选项控制图片位置
\centering
\includegraphics[width=1\textwidth]{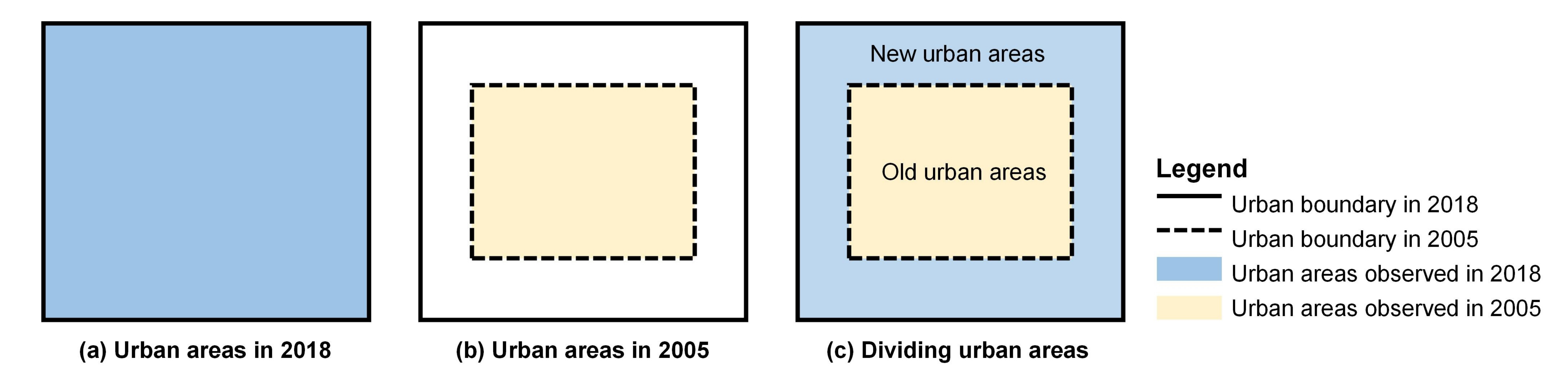}  
\caption{\textbf{The process of dividing urban areas}}
\label{fig:stream}
\end{figure}
\FloatBarrier % 确保图片在此之前出现

\subsection{Density of road network}
In this study, we harnessed global road network data from OpenStreetMap to assess morphological urban vitality globally using road network density as a metric. According to the visual interpretation and evaluation of satellite images, some researchers have proved that 83\% of the global road data of OSM in 2017 has been completed, and more than 40\% of countries have complete road networks \citep{BM2017}. The road network data, collected on February 1st, 2022, demonstrates consistent comparability across cities spanning various countries and regions. Using spatial analysis tools within the ArcGIS software, we processed this data to visualize the global distribution of road network density. Integrating it with global urban area data, we derived road network density figures for both old and new urban areas on a global scale.

\subsection{Density of points of interests}
To measure global functional urban vitality based on the density of points of interest (POIs), we collected global POIs data from OpenStreetMap, similar to the data source for road network density. The data collection date was February 1st, 2022. Researchers have proved that POI data in OSM after 2020 is more comprehensive, and there is a high correlation between the counting POIs mode and the statistics of Google Street View in several European cities \citep{PI2023}. Considering OSM provides the only publicly licensed data sources, the POIs data used in this paper comes from the OSM in 2022. Using this data and global urban area data, we calculated the density of POIs for old and new urban areas at a global scale.

\subsection{Density of road network}
Population density refers to the number of people per unit area. To assess global urban vitality from a social perspective, this study obtained 2020 population data worldwide from WorldPop \citep{TA2017}, which generates high-resolution open data on population distribution. The data were derived from a mosaiced global population dataset with a resolution of 1km, using 100m resolution counts \citep{LL2019}. By combining the global urban boundary data, the values of population density in old and new urban areas can be calculated. WorldPop's model is robust in identifying significant spatial relationships between census data and machine learning output, and excluding some rural population not found in built-up areas detected by satellite data \citep{PO2020} to improve accuracy. The model is transparent with its methods and data sources, integrating a variety of data to redistribute population counts across census or administrative units \citep{ST2015}, and allowing global comparisons over time. As this study focuses on urban areas, WorldPop is suitable as the data source.

\section{Results}\label{secIntro}
\subsection{Indicators between new and old urban areas indicate potential ghost cities in new urban areas}
To gain a deeper understanding of the concept of ghost cities, it is crucial to first compare several key indicators between new and old urban areas (Fig. 3). According to the definition of ghost city, the ghost city index of a city depends on the difference of density of road network, density of POIs, and population density between new and old urban areas. Based on the standard of the United Nations (\url{https://unstats.un.org/unsd/methodology/m49/}), the geographic regions are divided into the continents.

Globally, new urban areas distinctly exhibit lower POIs density compared to old ones. However, the difference in population density is relatively modest, with the variance in road network density being the most minimal. In these new urban areas, the road network density constitutes 42.51\%, the POIs density amounts to 19.45\%, and the population density reaches 38.60\%, when compared with those in older urban areas.

At the regional level, differences in road network density are generally minor, with Asia exhibiting the most substantial disparity. In this region, the road network density in new urban areas represents only 41.07\% of that in older urban settings. A notable instance is China, where extensive real estate developments in newer urban districts have led to relatively less dense road networks. POIs density differences are striking, both between new and old urban areas and across regions. Asia and North America display lower average POIs densities, while Europe exhibits the highest. In regions excluding Oceania, new urban areas have a POIs density below 20\% of older areas. Oceania, however, stands out with the smallest disparity, its comparative ratio being 32.72\%. In terms of population density, the gap is most pronounced in Asia, where new urban areas account for just 28.63\% of older areas. Conversely, Oceania showcases the least disparity, with a comparative ratio of 32.72\%.

% 插入图片
\begin{figure}[h!] % 使用 [!htbp] 选项控制图片位置
\centering
\includegraphics[width=0.92\textwidth]{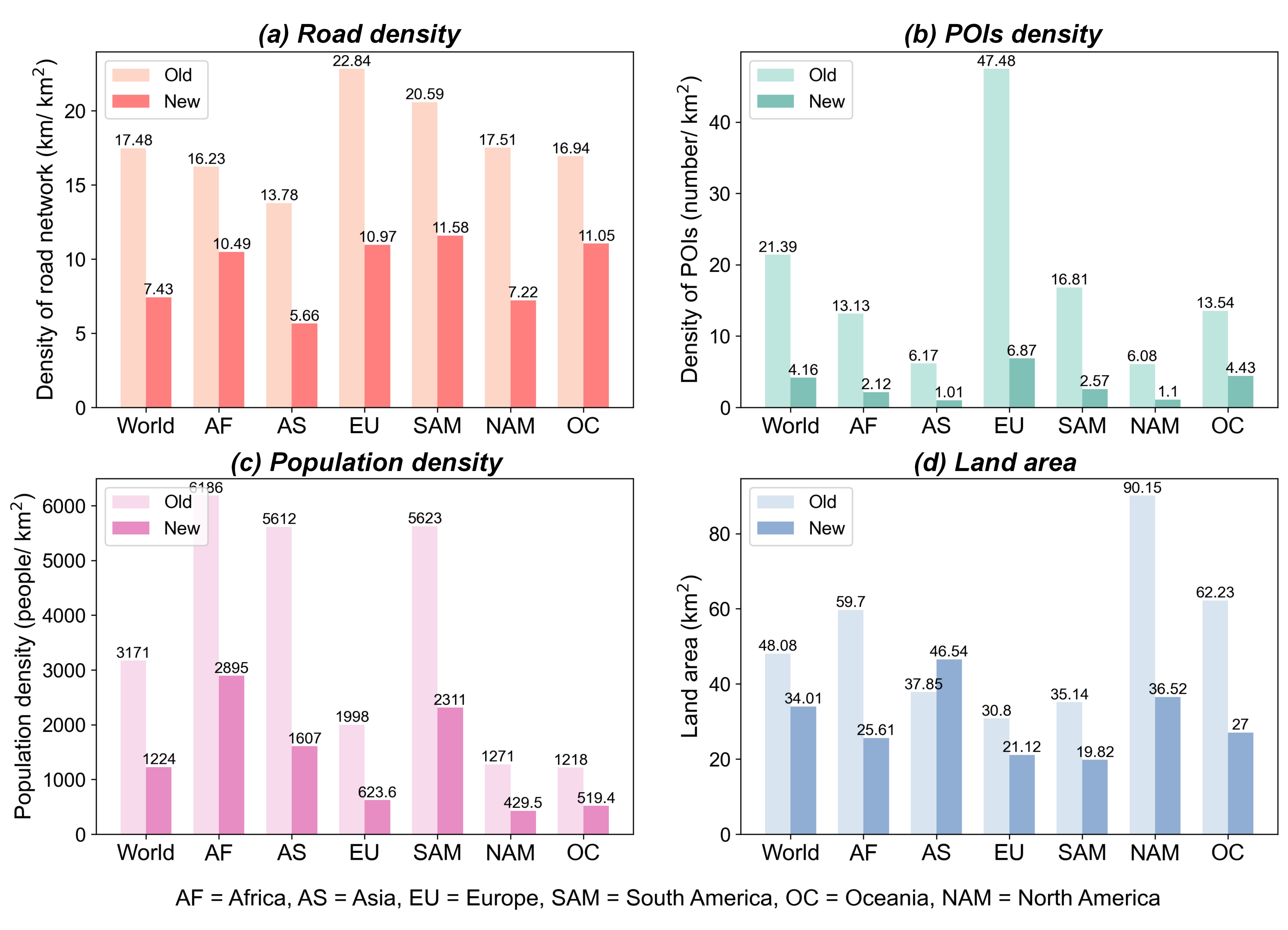}  
\caption{\textbf{ Indicators of global new and old urban areas by geographic regions in average.}}
\label{fig:stream}
\end{figure}
\FloatBarrier % 确保图片在此之前出现

The results at the national level are shown in Fig. 4. In terms of density of road network, the differences between countries and their new and old urban areas tend to be much smaller than other indicators. Conversely, the variance in POIs density between new and old urban areas across countries is notably more significant. In this dimension, most African countries exhibit lower levels, whereas many European countries demonstrate significantly higher levels. Regarding population density, a reverse trend is observed, with many European countries showing lower levels compared to African countries. The difference of the land area between new and old urban areas is relatively small for most countries, which is similar to density of road network. 

% 插入图片
\begin{figure}[h!] % 使用 [!htbp] 选项控制图片位置
\centering
\includegraphics[width=0.92\textwidth]{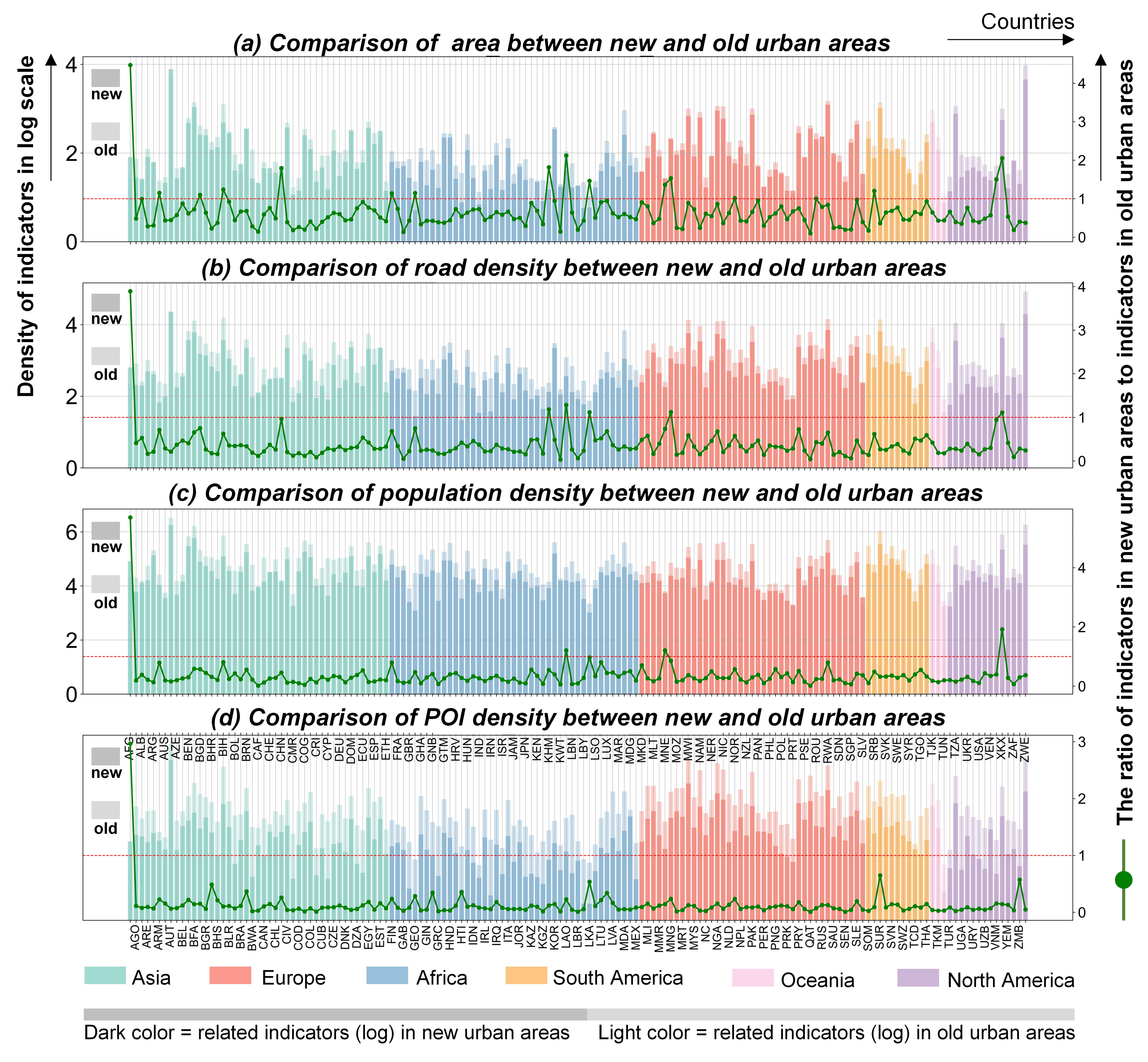}  
\caption{\textbf{Indicators for global ghost cities at the national average level.} Each bar represents a country, and different colors indicate different regions. ``old\_area'', ``new\_area'', ``roads\_old'', ``roads\_new'', ``old\_pop\_num'', ``new\_pop\_num'', ``poi\_old'', ``poi\_new'' represent the density of each indicator in new and old urban areas. All the real indicator values on the left are scaled by log 20 to facilitate visualization, and the ratio on the right is the real index values of the indicators in new urban areas to old urban areas. The last sub-picture gives the abbreviations of all countries (Supplementary Table 7).}
\label{fig:stream}
\end{figure}
\FloatBarrier % 确保图片在此之前出现

\subsection{The vitality of global urban areas at different levels}
To investigate the characteristics of global urban vitality, this study compares the vitality values of old and new urban areas at various levels, including city, national, regional, and global scales. Initially, the study examines the distribution patterns of urban vitality at the city level. Cities with higher vitality are mainly concentrated in Europe, East Asia, and South America as well as some tropical regions (Fig.3). Although the distribution patterns of vitality are broadly similar between new and old urban areas, it is notable that new urban areas tend to have fewer high vitality points compared to old urban areas. In more developed nations, such as the United States, the vitality differential between new and old urban areas is less pronounced than in developing nations, notably India and China. The vitality of new and old urban areas in Europe displays notable differences, suggesting a distinct development model from other developed countries, as indicated by the ‘ghost city index’ (Fig.7).

% 插入图片
\begin{figure}[h!] % 使用 [!htbp] 选项控制图片位置
\centering
\includegraphics[width=0.92\textwidth]{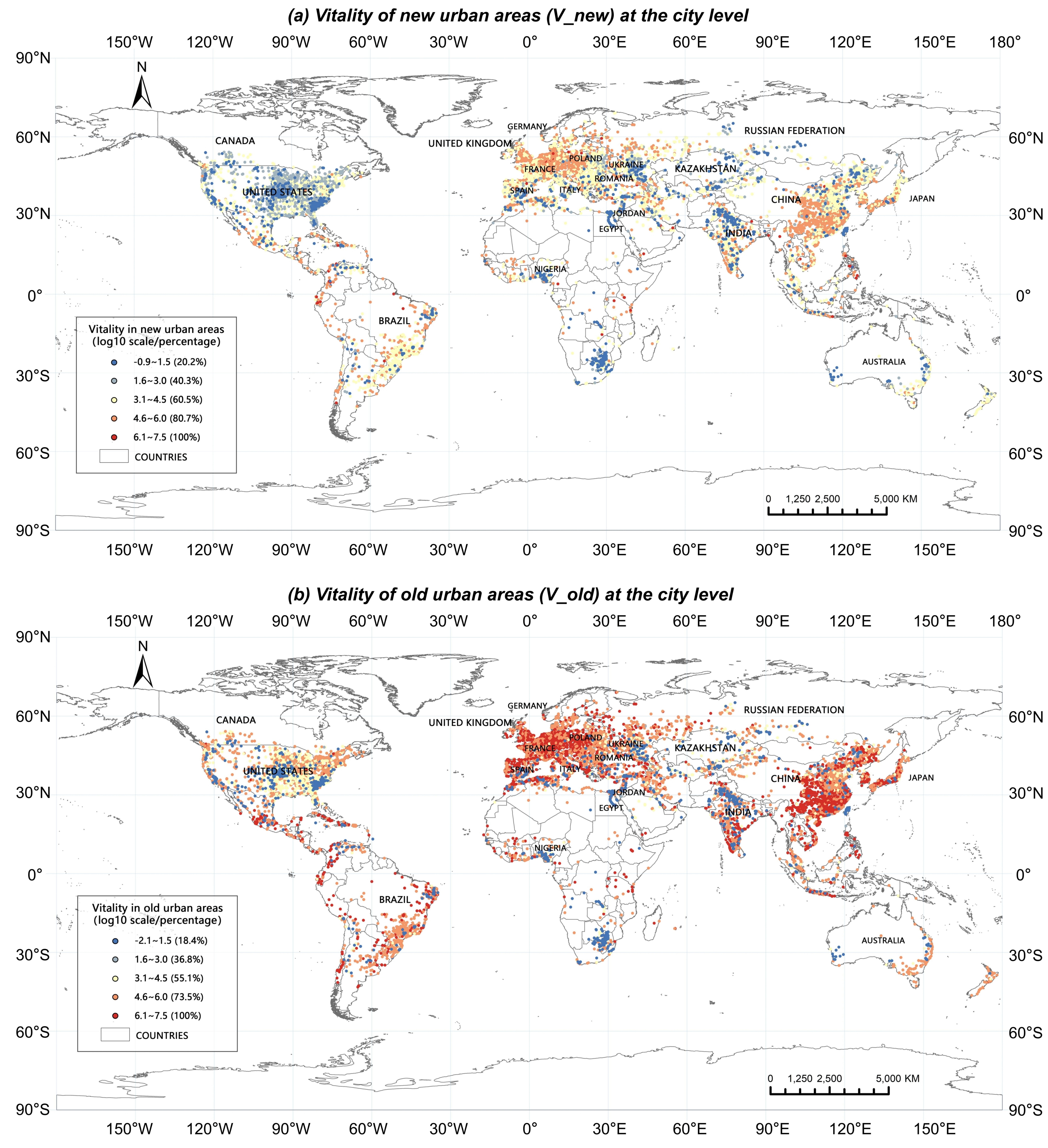}  
\caption{\textbf{The vitality of new and old urban areas for cities on the globe.} Each point represents a natural city. All the real index values on the left are scaled by log 10 to facilitate visualization. The ratio on the right side of the legend refers to the number of existing cities up to the interval to the total number of cities.}
\label{fig:stream}
\end{figure}
\FloatBarrier % 确保图片在此之前出现

Then the study analyzed the vitality of global urban areas at national, regional and global level (Fig 6). For each level, we recalculated indicators to obtain average vitality values for old and new urban areas. Most new urban areas have vitality accounting for less than 10\% of their older counterparts at both national and regional levels. Globally, the vitality of new urban areas is just 7.69\% of that in older areas. This significant disparity underscores the low vitality in newer regions. While it is evident that new urban areas generally exhibit much lower vitality than older ones, these findings reinforce the notion that assessing the vitality difference between new and old urban areas is an effective approach to identify the ghost cities on the globe.  

% 插入图片
\begin{figure}[h!] % 使用 [!htbp] 选项控制图片位置
\centering
\includegraphics[width=0.92\textwidth]{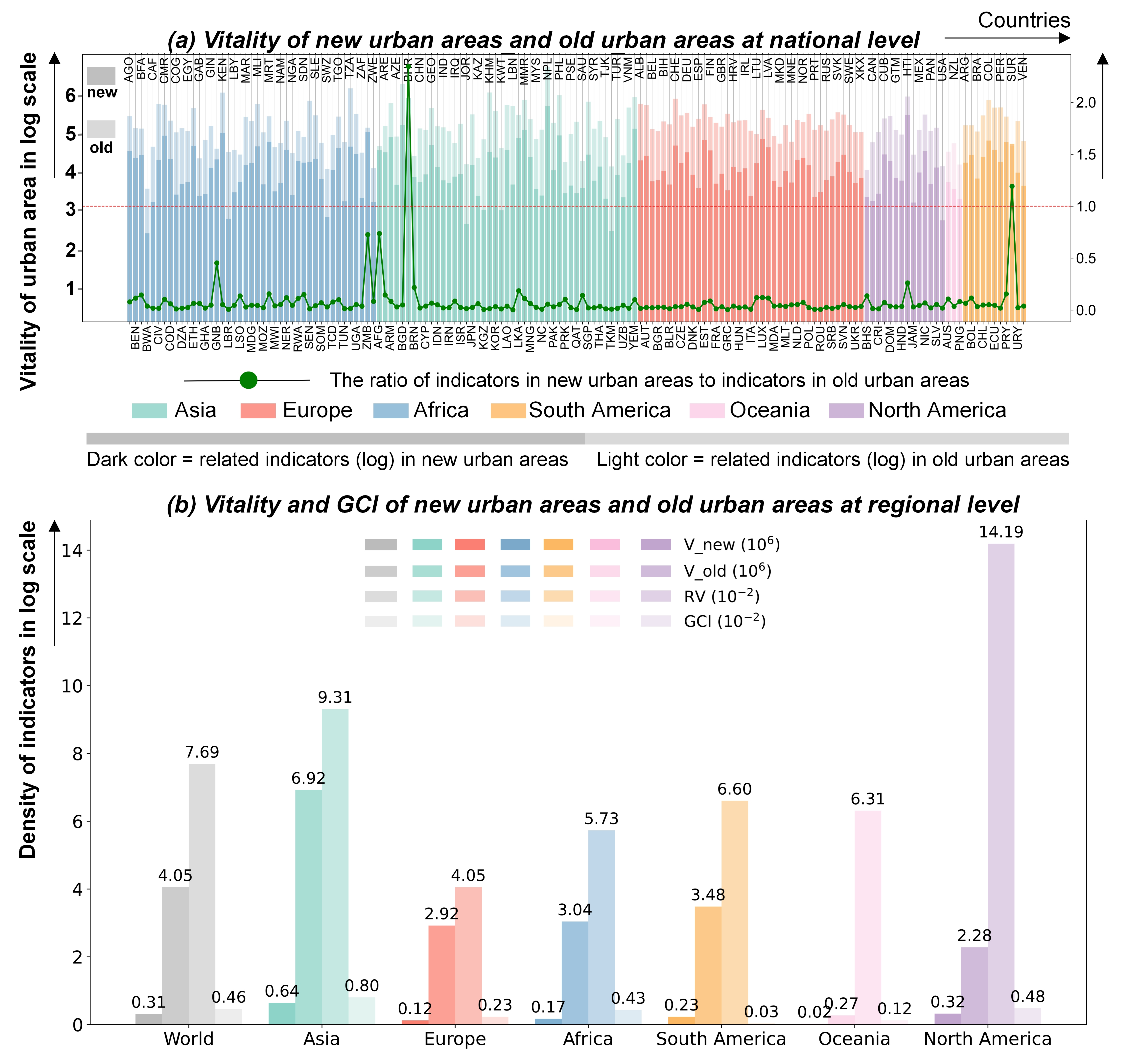}  
\caption{\textbf{Vitality of new urban areas and old urban areas at the national and regional level.} (a)Each bar represents a country, and different colors indicate different regions. All the real index values on the left are scaled by log 20 to facilitate visualization, and the ratio on the right means the real vitality index in new urban areas to that in old urban ares at national scale. (b)RV=Ratio of vitality between new and old urban areas.}
\label{fig:stream}
\end{figure}
\FloatBarrier % 确保图片在此之前出现

\subsection{Visualization of ghost city index on the globe at city level}
Fig. 7 maps each city's Ghost City Index (GCI). Cities with elevated GCI values are particularly prevalent in East and South Europe, Northeast China, and certain regions of the United States, underlining the global scope of the ghost city challenge. In a broader view, United States of America has a notable number of ghost cities, whereas Asia presents fewer. Meanwhile, Europe and Africa have relatively high counts and proportions of such cities.

Interestingly, some cities, while exhibiting high vitality in absolute terms, also maintain elevated GCI values. This indicates a marked disparity in vitality between new and old urban sectors (Fig.7). For instance, while cities in Europe and Asia demonstrate high vitality, the GCI in Northern Europe is significantly lower than in East and South Asia. A closer look at Fig. 7(b) reveals a near-normal distribution of GCI values across cities in Asia and Europe. In contrast, a majority of African cities possess a lower GCI, indicating a state of imbalanced urban development.

% 插入图片
\begin{figure}[h!] % 使用 [!htbp] 选项控制图片位置
\centering
\includegraphics[width=0.92\textwidth]{figures/figure_7.pdf}  
\caption{\textbf{The ghost city index (GCI) for cities and countries on the globe.} (b) Violin diagram shows the distribution shape, median, and quartile distribution of ghost city index (GCI) calculated by each country. GCI are scaled by log 20 to facilitate visualization.}
\label{fig:stream}
\end{figure}
\FloatBarrier % 确保图片在此之前出现

\subsection{Inferring ghost cities’ distribution on the globe}
The ranking of GCI values of global natural cities was utilized to infer the distribution of ghost cities, examining the number of cities and countries beneath various GCI thresholds, the proportion of identified ghost cities in all natural cities within each country, and ultimately setting the top 5\% (442 out of 8841 natural cities) as a threshold (see Fig.6) for visualization. The specific rankings can be accessed in the attached Supplementary Tables. We then focused on the top ten countries with the highest number of ghost cities, as depicted in Fig.6. Notably, the United States, China, and Italy lead this list, having the most significant count of ghost cities within their new urban areas. When comparing the proportion of ghost cities to the total number of natural cities within a country, Kazakhstan, Italy, Canada, and the United States stand out, as showcased in Fig.6.  

% 插入图片
\begin{figure}[!htbp] % 使用 [!htbp] 选项控制图片位置
\centering
\includegraphics[width=1\textwidth]{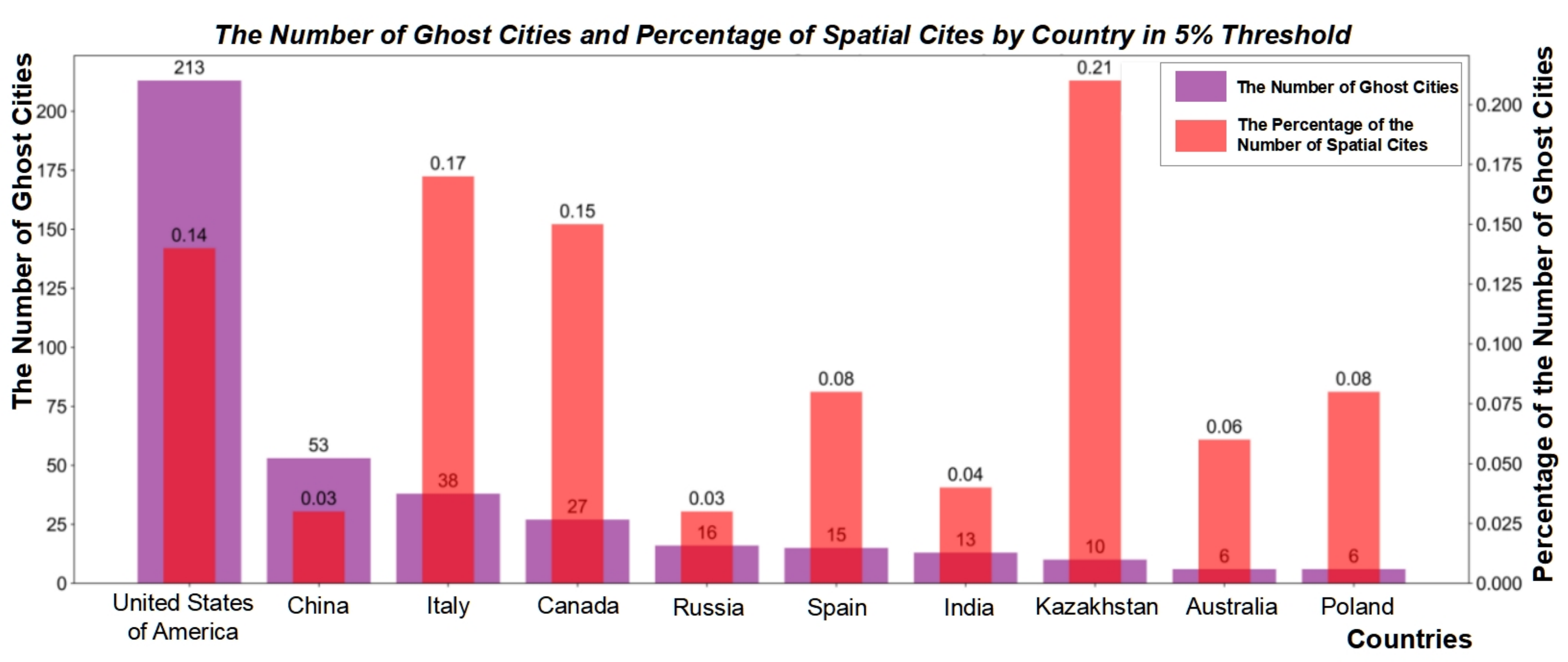}  
\caption{\textbf{Selection of ghost city threshold and the top ten countries with the largest number of ghost cities.}}
\label{fig:stream}
\end{figure}
\FloatBarrier % 确保图片在此之前出现

To illustrate our findings, we focus on the distribution of ghost cities in these countries ( Fig.9). Consistent with existing research, our findings also spotlight the United States and China as key regions of interest in the ghost city phenomenon. In the United States, ghost cities predominantly feature in the Midwest and eastern coastal areas. In contrast, China's ghost cities are primarily clustered in its northeast and northern regions. Overall, these findings suggest that while the phenomenon of ghost cities is present in multiple countries. However, the intensity and underlying reasons for these ghost cities differ across nations. 

% 插入图片
\begin{figure}[!htbp] % 使用 [!htbp] 选项控制图片位置
\centering
\includegraphics[width=1\textwidth]{figures/figure_9.pdf}  
\caption{\textbf{The ghost index (GCI) for the top ten countries with the largest number of ghost cities and a zoomed-in map for identified ghost cities. Country\_a\_b}}
\label{fig:stream}
\end{figure}
\FloatBarrier % 确保图片在此之前出现

% 插入图片
\begin{figure}[!htbp] % 使用 [!htbp] 选项控制图片位置
\centering
\includegraphics[width=1\textwidth]{figures/figure_10.pdf}  
\caption{\textbf{The ghost index (GCI) for the top ten countries with the largest number of ghost cities and a zoomed-in map for identified ghost cities. Country\_c\_d}}
\label{fig:stream}
\end{figure}
\FloatBarrier % 确保图片在此之前出现

% 插入图片
\begin{figure}[!htbp] % 使用 [!htbp] 选项控制图片位置
\centering
\includegraphics[width=1\textwidth]{figures/figure_11.pdf}  
\caption{\textbf{The ghost index (GCI) for the top ten countries with the largest number of ghost cities and a zoomed-in map for identified ghost cities. Country\_e\_f\_g}}
\label{fig:stream}
\end{figure}

\newpage
% 插入图片
\begin{figure}[!htbp] % 使用 [!htbp] 选项控制图片位置
\centering
\includegraphics[width=1\textwidth]{figures/figure_12.pdf}  
\caption{\textbf{The ghost index (GCI) for the top ten countries with the largest number of ghost cities and a zoomed-in map for identified ghost cities. Country\_h\_i\_j}}
\label{fig:stream}
\end{figure}

\clearpage
\subsection{Benchmarking our results from news media and related research}
To validate the inferred ghost cities in new urban areas, we compared our calculation results with news media and existing related rankings. First of all, we directly combine ‘ghost city’ with the names of all cities in the world (data comes from GADM), search on Google, and rank the number of search results of each city as the measure of the public's perception and media interest of ghost cities. Table 2 ranks the top 20 countries with the largest proportion of search result cities at different segments (Top 5\%-25\% cities) in ranked Google search results. We can find that they are highly consistent with the ghost cities identified by GCI at the national level (Fig 8), such as the United States, Canada, Italy, and Russia. They exist in different segments, and the high ranking of the United States’ search index further verifies the results of GCI. Overall, the search result indicates that our identification results for ghost cities align with the general perception of such cities worldwide. 

\begin{small} % 开始调整表格的整体字体大小
\begin{longtable}{|m{1cm}|m{2.4cm}|m{2.4cm}|m{2.4cm}|m{2.4cm}|m{2.4cm}|}
\caption{Top 20 countries with the most ghost cities in Google search results} \\
\hline
\textbf{Rank} & \textbf{Top 5\%} & \textbf{Top 10\%} & \textbf{Top 15\%} & \textbf{Top 20\%} & \textbf{Top 25\%} \\
\hline
\endfirsthead
\hline
\textbf{Rank} & \textbf{Top 5\%} & \textbf{Top 10\%} & \textbf{Top 15\%} & \textbf{Top 20\%} & \textbf{Top 25\%} \\
\hline
\endhead
1 & \textbf{*United States} & \textbf{*United States} & \textbf{Philippines} & \textbf{Philippines} & \textbf{Philippines} \\
  & 0.346 & 0.250 & 0.008 & 0.235 & 0.248 \\
\hline
2 & \textbf{*Canada} & \textbf{Philippines} & \textbf{*United States} & \textbf{*United States} & \textbf{*United States} \\
  & 0.222 & 0.177 & 0.006 & 0.147 & 0.120 \\
\hline
3 & \textbf{Philippines} & \textbf{*Canada} & \textbf{*Canada} & \textbf{*Canada} & \textbf{*Canada} \\
  & 0.132 & 0.161 & 0.005 & 0.129 & 0.120 \\
\hline
4 & \textbf{Germany} & \textbf{Germany} & \textbf{Germany} & \textbf{Germany} & \textbf{Germany} \\
  & 0.084 & 0.110 & 0.005 & 0.100 & 0.096 \\
\hline
5 & \textbf{United Kingdom} & \textbf{*Italy} & \textbf{*Italy} & \textbf{*Italy} & \textbf{*Italy} \\
  & 0.036 & 0.058 & 0.003 & 0.085 & 0.096 \\
\hline
6 & \textbf{Portugal} & \textbf{Portugal} & \textbf{Portugal} & \textbf{Thailand} & \textbf{Thailand} \\
  & 0.025 & 0.029 & 0.003 & 0.040 & 0.043 \\
\hline
7 & \textbf{*Russia} & \textbf{United Kingdom} & \textbf{Thailand} & \textbf{Portugal} & \textbf{Portugal} \\
  & 0.024 & 0.029 & 0.003 & 0.039 & 0.042 \\
\hline
8 & \textbf{Thailand} & \textbf{Thailand} & \textbf{United Kingdom} & \textbf{*India} & \textbf{*India} \\
  & 0.022 & 0.026 & 0.003 & 0.019 & 0.021 \\
\hline
9 & \textbf{*Italy} & \textbf{*Spain} & \textbf{*India} & \textbf{Switzerland} & \textbf{Indonesia} \\
  & 0.016 & 0.015 & 0.003 & 0.017 & 0.018 \\
\hline
10 & \textbf{*Austria} & \textbf{*India} & \textbf{Estonia} & \textbf{United Kingdom} & \textbf{Switzerland} \\
  & 0.008 & 0.014 & 0.002 & 0.017 & 0.017 \\
\hline
11 & \textbf{*India} & \textbf{*Russia} & \textbf{Switzerland} & \textbf{Estonia} & \textbf{Estonia} \\
  & 0.008 & 0.013 & 0.002 & 0.016 & 0.014 \\
\hline
12 & \textbf{Peru} & \textbf{Switzerland} & \textbf{*Spain} & \textbf{Indonesia} & \textbf{United Kingdom} \\
  & 0.008 & 0.013 & 0.002 & 0.013 & 0.014 \\
\hline
13 & \textbf{Estonia} & \textbf{Estonia} & \textbf{*Russia} & \textbf{*Poland} & \textbf{*Poland} \\
  & 0.008 & 0.012 & 0.002 & 0.013 & 0.013 \\
\hline
14 & \textbf{Panama} & \textbf{*Austria} & \textbf{Peru} & \textbf{Peru} & \textbf{Peru} \\
  & 0.007 & 0.011 & 0.002 & 0.012 & 0.012 \\
\hline
15 & \textbf{Switzerland} & \textbf{Peru} & \textbf{*Austria} & \textbf{*Spain} & \textbf{*Austria} \\
  & 0.007 & 0.010 & 0.001 & 0.011 & 0.010 \\
\hline
16 & \textbf{Chile} & \textbf{Costa Rica} & \textbf{*Poland} & \textbf{*Austria} & \textbf{*Spain} \\
  & 0.005 & 0.008 & 0.001 & 0.011 & 0.010 \\
\hline
17 & \textbf{Ecuador} & \textbf{France} & \textbf{Indonesia} & \textbf{*Russia} & \textbf{*Russia} \\
  & 0.005 & 0.007 & 0.001 & 0.010 & 0.009 \\
\hline
18 & \textbf{*Spain} & \textbf{*Poland} & \textbf{Costa Rica} & \textbf{Costa Rica} & \textbf{Ecuador} \\
  & 0.004 & 0.006 & 0.001 & 0.009 & 0.009 \\
\hline
19 & \textbf{*France} & \textbf{Panama} & \textbf{France} & \textbf{Ecuador} & \textbf{Costa Rica} \\
  & 0.004 & 0.005 & 0.001 & 0.008 & 0.008 \\
\hline
20 & \textbf{*Poland} & \textbf{Chile} & \textbf{Ecuador} & \textbf{France} & \textbf{*China} \\
  & 0.004 & 0.005 & 0.001 & 0.007 & 0.008 \\
\hline
\multicolumn{6}{|p{15cm}|}{\textbf{a} In order to keep the global unified administrative boundary division and acquire all cities’ names, we use the multilevel administrative boundary data from the Database of Global Administrative Areas (GADM, \url{https://gadm.org/}), and adopt level 3 as the retrieval combination mode (province/state + major cities + towns + ghost city/town). The ranking of ghost cities adopted by this study is the smallest division from the location of natural cities among the three-level administrative divisions in the world.} \\
\multicolumn{6}{|p{15cm}|}{\textbf{b} Different segmentation, such as 5\%, means that the top 5\% of all cities in Google search results are used as the threshold of ghost cities, and then the countries to which these cities belong are counted and ranked according to the proportion of ghost cities owned by countries to all natural cities in the country.} \\
\multicolumn{6}{|p{15cm}|}{\textbf{c} * represents the country in the 20 countries in this percentage grading that coincide with the countries in Fig 6.} \\
\hline
\end{longtable}
\end{small} % 结束字体调整

While several studies have depicted ghost cities using data such as population and night-time light, no one has quantitatively ranked the global ghost cities index according to our investigation, making it difficult to compare the rationality of our index with previous research on a global scale. To address this, we benchmarked our results with empirical research and news media at different regional scales, such as the Biaozhun ranking, \citet{SHI2020}, \citet{PK2019} and \citet{CP2020}, focusing on China and the United States (Supplementary Table 6). In terms of ghost cities in China, the Biaozhun ranking and the research of \citet{SHI2020} have some similarities compared with our results. We all agree that top ghost cities are mainly distributed in the northeast and northern China, including some cities like Wuzhong (Ningxia), Aletai (Xinjiang), Yichun (Xinjiang), Weihai (Shandong), and Qitaihe (Heilongjiang). As for ghost cities in the United States, the news reports of \citet{PK2019} and the 24/7 Wall St Ranking show many ghost cities in the United States consistent with our research results, both tending to locate ghost cities in the middle and northern regions of the United States. However, the news media and related research are more concerned about the vacancy of houses, so their results have limitations in application.

\subsection{Understanding the distribution of GCI from different perspectives}
Based on the derived 8841 natural cities, we calculated the population change and land expansion through the WorldPop data and understood the global distribution of GCI from the perspective of population change (Fig 10). Firstly, a total of 1825 natural cities have experienced population shrinkage and 7016 natural cities have experienced population expansion. Secondly, Fig 8 indicates that population growth and land expansion are often concentrated in large cities, especially in North America and Asia, showing the highest urban expansion speed. At the same time, compared with other regions, the population growth rate in South America and Africa is higher than the land expansion rate. Population growth faster than land expansion will lead to a high vitality index of new cities, which can also explain why the ghost city index in these areas is relatively low. On the contrary, the population expansion rate of a large number of North American cities is relatively low compared with the land expansion rate, which accounts for the existence of a large number of ghost cities in new urban areas in the United States.

% 插入图片
\begin{figure}[h!] % 使用 [!htbp] 选项控制图片位置
\centering
\includegraphics[width=1\textwidth]{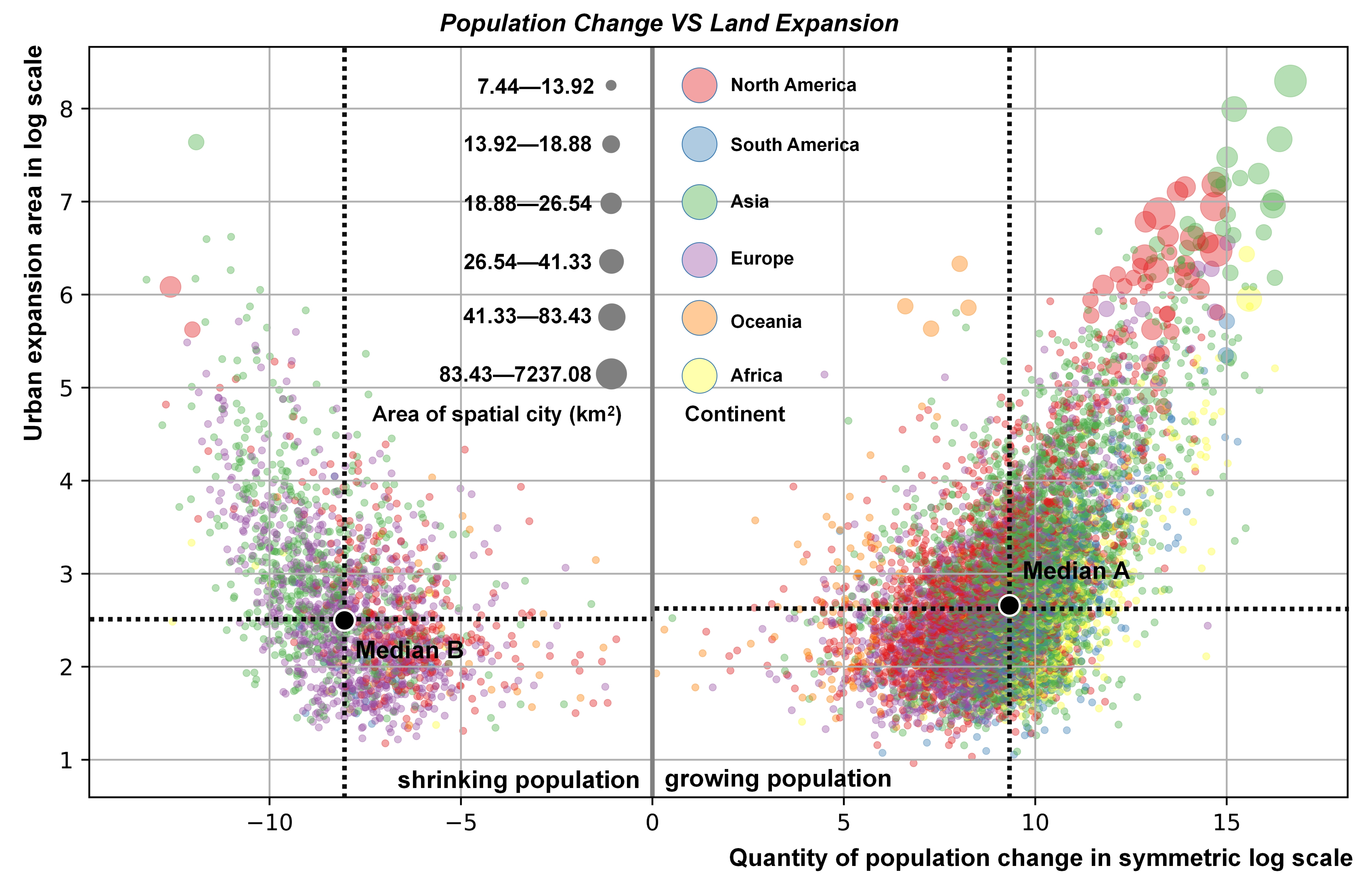}  
\caption{\textbf{Relationship between population change and land expansion in all the 8841 natural cities around the world.} The vertical axis at the zero point of the horizontal axis represents the relationship between population change and land use expansion under the two conditions respectively. Medians A and B represent the specific values of the median city of population expansion (contraction) and the median city of land expansion respectively. In order to facilitate visualization, symmetric logarithmic transformation is used for population change data and logarithmic transformation is used for land expansion data.}
\label{fig:stream}
\end{figure}

\section{Discussion and conclusion}
This paper presents a bottom-up method by utilizing publicly available open-access data to investigate the phenomenon of ghost cities in new urban areas across all natural cities worldwide. We claim that the conceptualization of a "ghost city" is characterized by the presence of expansive urban form, undeveloped urban functions, and very sparse human mobility/activity density \citep{JN2017}. For the first time, we attempt to rank ghost cities on the globe and provide an overall identification and verification of cities on the globe at different scales.

In the results of identifying ghost cities, we not only verify the known facts, such as the distribution of ghost cities confirmed by previous studies on local scales but also summarize the differences in the development of new urban areas in different countries and regions. First, countries and regions in different urbanization stages have different characteristics of urban vitality and ghost cities. Although the results confirmed again that new urban areas have a lower vitality level than old urban areas, the difference in the ghost city index among Africa, Asia, and Europe shows the different urbanization stages of these regions. At the stage of vast urban expansion in Asia, many new urban areas are built up without enough infrastructure, facilities, and population, resulting in a higher ghost city index. At the later period of urbanization like in some European countries, especially in western and northern Europe, the contrary is the case.

Second, for all the cities worldwide, we can find that the high ghost city index mainly comes from the lack of the functional dimension of the new urban areas. The road density usually shows a small gap compared with other indicators in the new and old urban areas, which is easy to understand because the expansion of urban built-up areas is often ahead of the migration of population and the supplement of facilities policies, especially in some developing countries. At the same time, the difference in activity density is similar to the road density at the global level, far less than the difference in functional infrastructure. However, at the regional level, it can be found that there are great differences in the development of new urban areas in different countries and regions. The high index of Asian ghost cities is the joint result of the interaction of three dimensions, which probably comes from the fact that compared with other regions, Asian cities have the fastest expansion. Africa, which is also a gathering place of developing countries, shows different characteristics from Asia in the ghost city index, and more comes from population activities and functional infrastructure.

Third, the findings of this paper highlight the widespread distribution of ghost cities as a severe global problem, particularly in North America, Asia, and parts of Europe. The results of our identification on a global scale show that the distribution of ghost cities is concentrated in the United States, China, Italy, and other major countries. Unlike what we expected, the United States, as a developed country, has a large number of ghost cities compared with other countries. On the one hand, we speculate that many developed countries have been affected by the cyclical economic crisis, and some so-called ghost cities may be related to the long-term impact of the 2008 crisis, especially in the United States. Although China does not have a high proportion of ghost cities, due to rapid urbanization, it still has many ghost cities. On the local scale, our ghost cities are located in the northeast of China and the east coast of the United States, which is consistent with the previous research findings and news media \citep{JN2017, SHI2020}. This reveals the significance of spatial planning and development strategies in coping with ghost cities and building sustainable cities. Our validation results prove the rationality of this method, in which the search rankings for Chinese ghost cities are notably lower across the segments. This could be attributed to the limited popularity of Google searches in China.

According to the results, policy implications can be provided in the following aspects. First, since the high ghost city index is mainly caused by the functional dimension of new urban areas at the global scale, functions in new urban areas of many developing countries need significant improvement. The lower density of POIs reflects fewer functions, including amenities, public services, and jobs. More functions can be the key to mitigating ghost cities on the global scale. Second, the policy-making of urbanization strategies should consider the urbanization stage. For low-income countries like some African countries, it could be better to make an evaluation of the development situation in old urban areas and keep a proper construction size of new urban areas. Third, spatial planning in new urban areas should consider proper population density and optimize land use configuration, especially in developing countries in Asia. Denser roads should be taken into account to improve spatial accessibility and urban vitality.

This study shows that identifying ghost cities on the globe and their spatial differences is of great significance for understanding the urbanization process and building a sustainable city. Some local scale studies have discussed the definition and recognition of ghost cities, but they have not yet formed a globally recognized recognition framework. Nevertheless, we believe that our bottom-up approach is a straightforward and sustainable framework for evaluating ghost cities. Different from the recent research, we pay more attention to the comparability and easy use of the identification of ghost cities in the world. By dividing the boundaries of natural cities based on the time series and using publicly available open-access data, we overcome the limitation that previous studies can only focus on the identification of ghost cities on a certain local scale. This comparison based on its urban spatial development shows the advantages of crossing geographical spatial heterogeneity. While this comparison based on urban spatial development transcends geographical heterogeneity, the determination of the GCI threshold for defining a ghost city remains somewhat subjective. Future research should delve deeper into the causes, stages, and critical values of the ghost city phenomenon. Additionally, given the data limitations like OSM at the global level, the analysis is expected to be further deepened with the data environment continues to evolve.

\section*{Code availability}
All reference data come from open databases, and their download channels have been shown in the method section (refer to Table 2). The primary datasets and codes of this paper have been uploaded to “BCL-Global-Ghost-Cities, Mendeley Data, V1, doi: 10.17632/ypj9s32tn6.1”, which describes all the data generated in this paper. This includes the spatial location of 8842 natural cities calculated in this paper, the attributes of new and old urban areas and the ghost city index.

\section*{Author contributions statement}
YZ: Writing - original draft, Software, Visualization, Validation; TT:Writing - original draft, Formal Analysis, Investigation; YL: Conceptualization, Funding acquisition, Project Administration, Supervision.

\section*{Acknowledgements} 
Thanks to Master Qingxiang Meng for his contributions to the original data calculation.

\section*{Competing interests}
All authors declare no financial or non-financial competing interests.

\newpage % 或者 \clearpage

\clearpage % 或者 \newpage
\section*{Supplemental Material}
Overview: Appendices can be requested from the author or wait for the official publication of the paper.\\
\textbf{Supplementary Notes}\\
Supplementary note 1. Data cleaning on a city scale.\\
\textbf{Supplementary Tables}\\
Supplementary Table 1. Area ratio and attributions between GNUA and GOUA.\\
Supplementary Table 2. Area ratio (Ratio of GNUA to GOUA) <0.15 and GNUA< 4 km2.\\
Supplementary Table 3. Area ratio (Ratio of GNUA to GOUA) >20 and GOUA < 4 km2.\\
Supplementary Table 4. The global rankings of ghost cities.\\
Supplementary Table 5. The rankings of ghost cities in China.\\
Supplementary Table 6. The rankings of ghost cities in United States.\\
Supplementary Table 7. The full names of countries involved in this article corresponding to the abbreviations.\\
\textbf{Supplementary Figures}\\
Supplementary Figure 1. The number of cities and Sum GNUA under different new urban area thresholds when ratio<0.15.\\
Supplementary Figure 2. The number of cities and Sum GOUA under different old urban area thresholds when ratio>20.\\

\end{document}